\documentclass[useAMS,usenatbib]{mn2e}

\usepackage{graphicx}	
\usepackage{amsmath}	
\usepackage{amssymb}	

\title[No sign of IGCs in the Local Group]{No sign (yet) of intergalactic globular clusters in the Local Group}

\author[Mackey, Beasley \& Leaman]{A. D. Mackey$^{1}$\thanks{E-mail: dougal.mackey@anu.edu.au}, M. A. Beasley$^{2,3}$ and R. Leaman$^{4}$\\
$^1$Research School of Astronomy and Astrophysics, Australian National University, Canberra, ACT 2611, Australia\\
$^2$Instituto de Astrof\'isica de Canarias, Calle Via Lactea, La Laguna, Tenerife, Spain\\
$^3$University of La Laguna, Avda. Astrof\'isico Fco. S\'anchez, La Laguna, Tenerife, Spain\\
$^4$Max-Planck Institut f\"{u}r Astronomie, K\"{o}nigstuhl 17, D-69117 Heidelberg, Germany
}

\date{Draft version \today.}

\pubyear{2016}

\begin{document}
\label{firstpage}
\pagerange{\pageref{firstpage}--\pageref{lastpage}}
\maketitle

\begin{abstract}
We present Gemini/GMOS imaging of twelve candidate intergalactic globular clusters (IGCs) in the Local Group, identified in a recent survey of the SDSS footprint by \citet{dtzz:15}. Our image quality is sufficiently high, at $\sim 0.4\arcsec - 0.7\arcsec$, that we are able to unambiguously classify all twelve targets as distant galaxies. To reinforce this conclusion we use GMOS images of globular clusters in the M31 halo, taken under very similar conditions, to show that any genuine clusters in the putative IGC sample would be straightforward to distinguish. Based on the stated sensitivity of the \citet{dtzz:15} search algorithm, we conclude that there cannot be a significant number of IGCs with $M_V \le -6$ lying unseen in the SDSS area if their properties mirror those of globular clusters in the outskirts of M31 -- even a population of $4$ would have only a $\approx 1\%$ chance of non-detection.
\end{abstract}




\section{Introduction}
Globular clusters are observed in many Local Group galaxies -- in abundance in the haloes of the Milky Way and M31, and in smaller numbers in roughly a dozen dwarf galaxies spanning a variety of morphological types. However, in the Local Group no examples of intergalactic globular clusters (IGCs) are known. This is in stark contrast to the situation in denser environments, such as large galaxy clusters, where substantial populations of IGCs are seen \citep[e.g.,][]{gregg:09,peng:11}. It is unclear whether the local dearth of IGCs is a consequence of observational bias (due to a lack of dedicated searches, and, until quite recently, sufficiently deep and uniform all-sky imaging) or whether it reflects an intrinsic scarcity of such objects.

A thorough review of the reasons why a population of Local Group IGCs might plausibly be expected has recently been presented by \citet{dtzz:15}. They argue that there are two possible formation channels. The first posits that globular clusters are formed in galaxies, but that strong galaxy-galaxy interactions might subsequently lead to some fraction becoming unbound. This is thought to be the origin of the significant IGC populations observed in galaxy clusters, where close encounters between galaxies frequently lead to tidal disruption or stripping \citep[e.g.,][]{west:11,samsing:15}. It is well known that galaxy interactions, mergers and accretions have occurred in the Local Group. Numerous stellar streams and overdensities are observed in the halo of the Milky Way \citep[e.g.,][]{belokurov:06,martin:14,grillmair:16}, and indeed the Sagittarius dwarf is presently in the process of being accreted \citep{ibata:94}. The stellar stream resulting from the tidal destruction of Sagittarius may be traced across the entire sky \citep[e.g.,][]{majewski:03}, and this debris includes a number of globular clusters that were once hosted by the dwarf \citep{bellazzini:03,law:10}. In M31 an abundance of halo substructure is seen out to radii beyond $100$ kpc in projection \citep{mcconnachie:09,ibata:14}, and many remote globular clusters are associated, both spatially and kinematically, with these features \citep{mackey:10a,mackey:13,mackey:14,veljanoski:13,veljanoski:14}.

The largest stream visible in the M31 halo is the Giant Stellar Stream, a signature of the most significant recent accretion event in the Local Group. This event involved an early-type progenitor which experienced an energetic, near head-on collision with M31 $\sim 1-2$ Gyr ago; prior to its disruption this system was likely the fourth or fifth most massive galaxy in the Local Group \citep{fardal:13}. Debris from this encounter can be traced to at least $100$ kpc from the M31 centre. Moreover, there is strong evidence that the third most-massive Local Group galaxy, M33, has recently interacted with M31, as has the dwarf elliptical satellite NGC 147. In both cases these interactions have been sufficiently strong so as to draw substantial stellar material from the inner parts of these galaxies into extended tidal tails \citep{mcconnachie:10,crnojevic:14}. In the Milky Way sub-group, recent observations of the periphery of the Large Magellanic Cloud (LMC) have revealed a $\sim 10$ kpc stellar substructure \citep{mackey:16} that may be a result of tidal stripping by the Milky Way, or repeated close interactions between the LMC and its smaller neighbour the Small Magellanic Cloud \citep[e.g.,][]{besla:16}.

While interactions between Local Group galaxies appear common, the dispersion of barycentric velocities is of order $\approx 50$ km$\,$s$^{-1}$ \citep[e.g.,][]{mcconnachie:12} such that the typical collision energy is likely much lower than in a dense galaxy cluster. Hence it is plausible that IGCs arising due to close encounters between Local Group galaxies are quite rare.

The second, more speculative, formation channel advanced by \citet{dtzz:15} is that IGCs might form {\it in situ}, within their own individual dark matter haloes \citep{peebles:84}.  While no good evidence for dark matter in globular clusters has yet been observed \citep[e.g.,][]{lane:10,ibata:13}, this hypothesis has not been definitively ruled out. Perhaps the only place one might find ``pristine'' globular clusters that formed in this way would be in intergalactic space; it is thought that clusters entering the dark matter halo of a large galaxy would have any of their own dark matter quickly stripped \citep[e.g.,][]{mashchenko:05}. The most isolated globular cluster known in the Local Group is MGC1, which sits $\approx 200$ kpc from the centre of M31 \citep{mackey:10b}. Nonetheless, this is still well within the expected virial radius of the system ($\sim 300$ kpc), and \citet{conroy:11} showed that the observed radial density profile of MGC1 precludes that this cluster resides within a dark matter halo of mass $\ga 10^6 M_\odot$.

Motivated by these questions, \citet{dtzz:15} recently conducted the first systematic large-area search for Local Group IGCs.  The basis of their survey is the Sloan Digital Sky Survey (SDSS) Galaxy Catalog, which spans $\sim 14\,500$ deg$^2$, or roughly one third of the entire sky.  By combining the SDSS optical data with infrared photometry from the {\it Wide-field Infrared Survey Explorer} ({\it WISE}) satellite, and ultraviolet measurements from the {\it Galaxy Evolution Explorer} ({\it GALEX}) satellite, they attempt to select objects with spectral energy distributions matching those observed for globular clusters in Local Group galaxies. This technique is demonstrably successful -- \citet{dtzz:15} present the discovery of $22$ clusters in the halo of M31 \citep[see also][]{dtzz:13,dtzz:14}, many of which have independent verification from the {\it Pan-Andromeda Archaeological Survey} \citep{huxor:14}. In addition to these objects, \citet{dtzz:15} list another $12$ candidates that pass their selection criteria but lie well away from M31 and all other Local Group galaxies. As such, they identify these objects as possible Local Group IGCs. 

In this paper, we present high quality ground-based imaging of these $12$ objects with the aim of assessing whether or not any of them are {\it bona fide} globular clusters.

\section{Observations and data reduction}
\label{s:data}
We obtained snapshot images of the $12$ candidate IGCs listed by \citet{dtzz:15} with the Gemini Multi-Object Spectrograph (GMOS) at the Gemini North telescope on Mauna Kea, Hawaii.  The observations were carried out in queue mode as program GN-2015B-Q-17 (PI: Mackey), between late July and early November 2015. The data were collected during clear conditions and under excellent seeing ($\sim 0.4\arcsec - 0.7\arcsec$). Table \ref{t:log} presents the observing log and the full list of targets.

\begin{table}
\centering
\caption{Log of observations.}
\begin{tabular}{@{}llllc}
\hline \hline
Target & RA & Dec & Date & Image \\
Name & (J2000) & (J2000) & Observed & Quality \\
\hline
dTZZ-C01 & $00\,54\,27.3$ & $+04\,11\,01.4$ & $2015$ Jul $26$ & $0.41\arcsec$ \\
dTZZ-C02 & $01\,09\,22.7$ & $-05\,54\,57.5$ & $2015$ Jul $28$ & $0.40\arcsec$ \\
dTZZ-C03 & $02\,05\,30.4$ & $+06\,46\,41.1$ & $2015$ Aug $27$ & $0.44\arcsec$ \\
dTZZ-C04 & $02\,30\,35.9$ & $+46\,19\,11.9$ & $2015$ Aug $20$ & $0.58\arcsec$ \\
dTZZ-C05 & $06\,48\,36.5$ & $-18\,23\,19.7$ & $2015$ Oct $08$ & $0.56\arcsec$ \\
dTZZ-C06 & $08\,03\,29.3$ & $+13\,04\,38.3$ & $2015$ Oct $08$ & $0.49\arcsec$ \\
dTZZ-C07 & $10\,09\,37.0$ & $+61\,15\,58.4$ & $2015$ Nov $07$ & $0.55\arcsec$ \\
dTZZ-C08 & $15\,54\,37.3$ & $+12\,55\,13.4$ & $2015$ Aug $02$ & $0.39\arcsec$ \\
dTZZ-C09 & $16\,40\,17.9$ & $+54\,58\,05.6$ & $2015$ Jul $26$ & $0.51\arcsec$ \\
dTZZ-C10 & $17\,03\,44.7$ & $+38\,47\,50.2$ & $2015$ Jul $26$ & $0.49\arcsec$ \\
dTZZ-C11 & $20\,55\,18.0$ & $+54\,42\,46.2$ & $2015$ Oct $08$ & $0.70\arcsec$ \\
dTZZ-C12 & $22\,54\,47.4$ & $+17\,26\,21.4$ & $2015$ Jul $26$ & $0.39\arcsec$ \\
\hline
\label{t:log}
\end{tabular}
\end{table}

For a given object we obtained two frames of exposure duration $145$s each, with a $5\arcsec$ dither to fill in the GMOS inter-CCD gaps. All imaging was conducted through the GMOS $i^\prime$ filter.  We reduced the data using standard procedures in the {\sc gmos} software package in {\sc iraf}.  Bias and flat-field images were applied with the {\sc gireduce} task, the three CCD frames in a given exposure were mosaicked into a single frame with {\sc gmosaic}, and then the two frames for a given object were stacked together using {\sc imcoadd}.  

\section{Results}
\begin{figure*}
\begin{center}
\includegraphics[width=158mm]{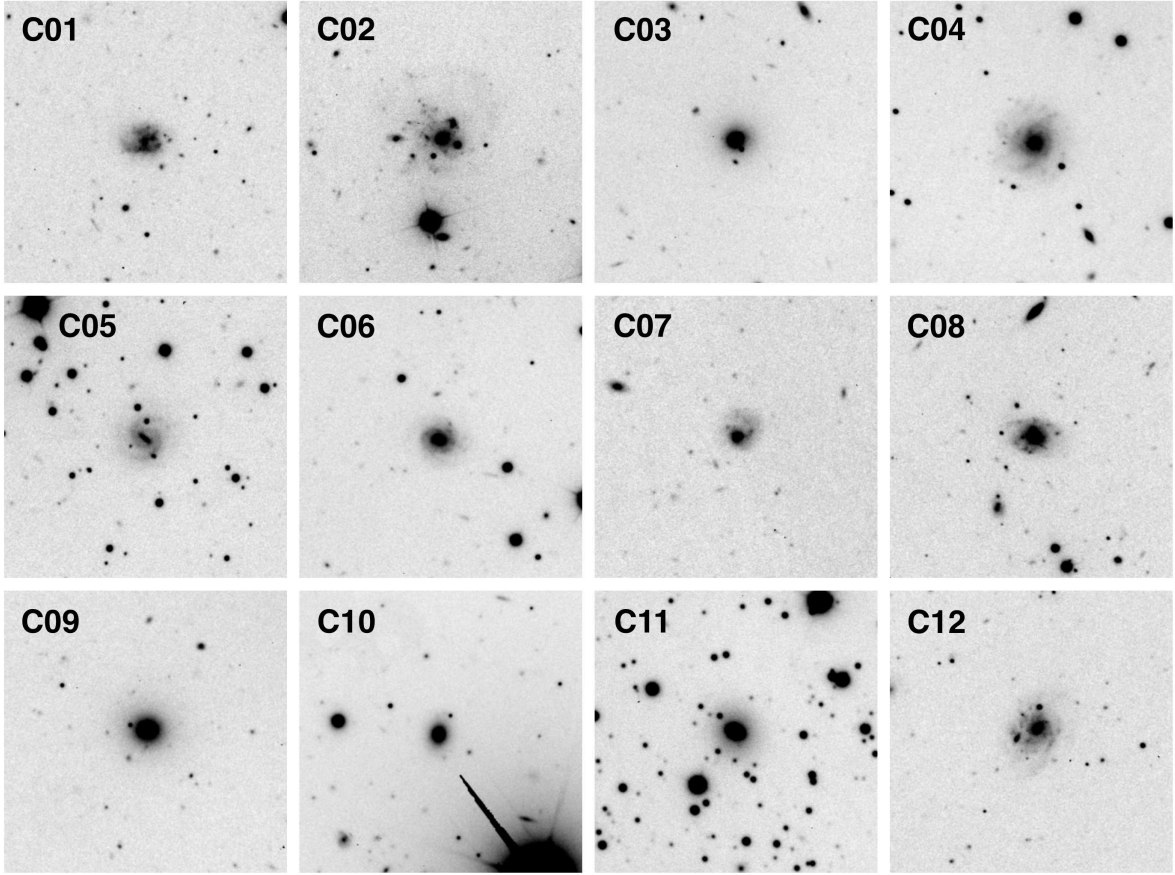}
\end{center}
\caption{GMOS $i^\prime$-band images of the $12$ candidate IGCs. Each thumbnail is $1\arcmin \times 1\arcmin$ and oriented such that north is up and east to the left.
\label{f:images}}
\end{figure*}

Images of the targets are shown in Figure \ref{f:images}.  These are $1\arcmin \times 1\arcmin$ cut-outs from the final reduced GMOS frames. It is evident that all $12$ of the IGC candidates are distant galaxies. Six exhibit distinct spiral arms (C02, C04, C05, C06, C07, and C12), while two have a more irregular morphology (C01 and C08). The remaining four (C03, C09, C10, and C11) appear to be early type galaxies. 

\begin{figure*}
\begin{center}
\includegraphics[width=158mm]{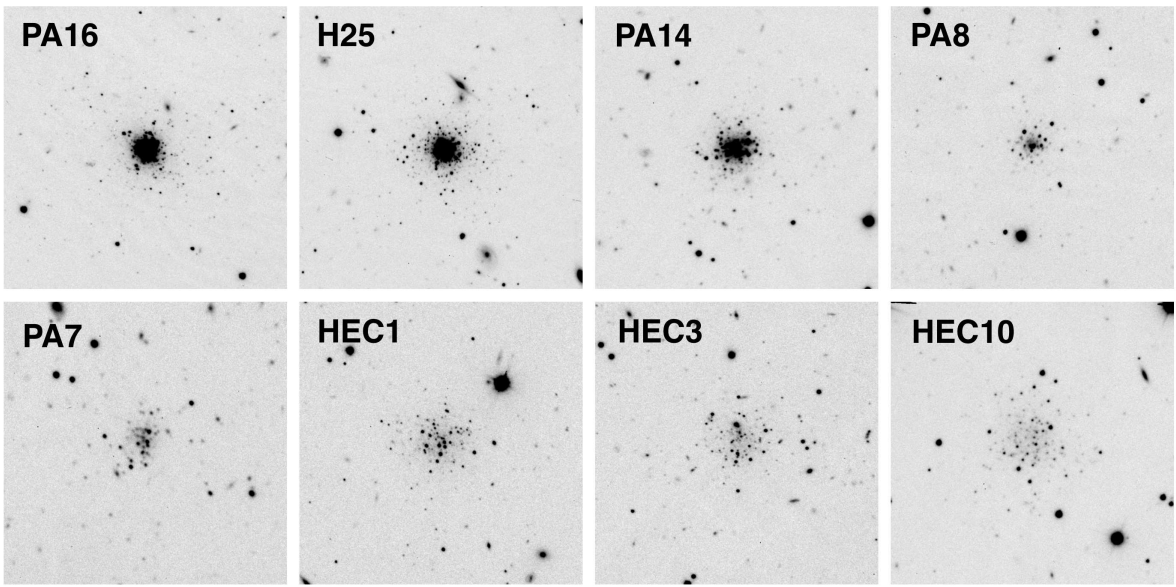}
\end{center}
\caption{GMOS $i^\prime$-band images of $8$ globular clusters in the outer M31 halo. These span luminosities of $-8.5 \la M_V \la -5.0$ and sizes $3 \la r_h \la 25$ pc, and sit at roughly the M31 distance of $780$ kpc. As before, each thumbnail is $1\arcmin \times 1\arcmin$ and oriented such that north is up and east to the left.
\label{f:clusters}}
\end{figure*}

To provide insight on what we might expect IGCs in the Local Group to look like in this type of image, Figure \ref{f:clusters} shows examples of globular clusters in the outer halo of M31 observed with GMOS through the $i^\prime$ filter as part of programs GN-2008B-Q-22 and GN-2014B-Q-26 (PI: Mackey). Atmospheric conditions were very similar to those under which our IGC images in Figure \ref{f:images} were obtained; seeing was in the range $\sim 0.3\arcsec - 0.6\arcsec$. Long sequences of $g^\prime$ and $i^\prime$-band images were taken in an effort to construct deep colour-magnitude diagrams for these clusters \citep[as in][]{mackey:13}; to ensure a fair comparison to our data for the IGC candidates, we selected and stacked only two random $i^\prime$-band images for each object in Figure \ref{f:clusters}.  The total exposure times are longer than for the IGC candidates, but not by more than a factor of two.

The clusters were chosen from the catalogues of \citet{huxor:08,huxor:14} to span a representative range in luminosity ($-8.5 \la M_V \la -5.0$) and size ($3 \la r_h \la 25$ pc).  The mean distance of the sample is approximately that of M31 -- i.e., $780$ kpc, corresponding to a distance modulus of $24.46$ \citep{conn:12}. However, because these objects sit in the outskirts of the M31 halo at projected radii in the range $45 \la R_{\rm p} \la 90$ kpc, there could plausibly be line-of-sight distance variations from object to object of order $\pm 100$ kpc about this value \citep[see, e.g.,][]{mackey:10b}. While the ``edge'' of the Local Group is an ill-defined concept, a reasonable estimate would be that it is not much more than $\approx 1$ Mpc from the midpoint of the vector connecting the Milky Way and M31 \citep[e.g.,][]{mcconnachie:12}. Thus these clusters sit at distances comparable to those that might be expected for the IGC candidates in the case where any of these was a real cluster in the Local Group -- recall that most fall well away from the direction of M31 on the sky.

Irrespective of their luminosity or physical size, all the globular clusters shown in Figure \ref{f:clusters} exhibit resolved stars.  In the case of the more compact clusters, these surround an unresolved core possessing an irregular appearance.  None of the clusters displays anything akin to a feature that could be mistaken for a spiral arm.  Similarly, none of them exhibits a diffuse, unresolved component similar to the irregular morphology of C01 or C08, or the much smoother morphology of C03, C09, C10, and C11.

On the basis of this simple comparison, we are confident that none of the objects shown in Figure \ref{f:images} is a star cluster.

\section{Discussion \& Conclusions}
\label{s:discussion}
The fact that there are no globular clusters amongst the $12$ candidates from \citet{dtzz:15} suggests that IGCs in the Local Group are intrinsically rare, although it is difficult to quantitatively assess the meaning of ``rare'' in this instance. \citet{dtzz:15} provide the following information about the efficiency, and limitations, of their detection algorithm:
\begin{itemize}
\item{The search area spans roughly one third of the sky, although a few regions (e.g., towards the Galactic plane) are affected by extinction and crowding.}
\item{They are only sensitive to clusters more luminous than $M_V \sim -6$ out to the edge of the Local Group (i.e., $\approx 1$ Mpc from the midpoint of the vector connecting the Milky Way and M31).}
\item{They are only sensitive to relatively compact clusters; objects like some of those in Figure \ref{f:clusters} that might be mostly resolved into stars in SDSS images do not appear in their base catalogue.}
\item{By passing data for known Local Group globular clusters through their selection criteria, they show that their detection completeness is $\approx 85\%$ for objects in the range of size and luminosity to which the algorithm is sensitive.} 
\end{itemize}
Given these constraints, we can ask how likely it is that a population of IGCs falls within the SDSS footprint but was not detected by the \citet{dtzz:15} search procedure.  To proceed, we make the assumption that the putative IGCs have the same luminosity function and size distribution as globular clusters observed in the outer halo of M31 at projected radii in the range $25 \le R_{\rm p} \le 150$ pc. We choose M31 because (i) it has {\it many} more globular clusters at such radii than does the Milky Way, so the population statistics are sounder; and (ii) perhaps surprisingly, the census of remote globular clusters is likely more complete in M31 than in the Milky Way -- as evidenced by the continuing discovery of moderately luminous clusters in the Milky Way periphery \citep[e.g.,][]{belokurov:14,laevens:14,laevens:15}.

From \citet{huxor:14} we see that, for M31 clusters with $R_{\rm p} \ge 25$ kpc, the distribution of sizes depends weakly on luminosity. We assume, conservatively, that the \citet{dtzz:15} search algorithm is sensitive to objects with $r_h \le 10$ pc, but does not see objects with sizes larger than this. At luminosities brighter than $M_V \approx -7.5$, all clusters in the M31 sample have $r_h \le 10$ pc; however, for luminosites in the range $-7.5 \le M_V \le -6$, approximately $40\%$ of clusters have $r_h > 10$ pc. Out of the clusters with $M_V \le -6$, half have $M_V \le -7.5$, and half have $-7.5 \le M_V \le -6$.

Armed with this information, we see that the chance that a cluster is detectable (i.e., the probability that it has $r_h \le 10$ pc given that its luminosity is $M_V \le -6$) is $0.8$. Applying the $85\%$ success rate stated by \citet{dtzz:15} then implies that for any given cluster the total chance of detection is $0.68$. Hence, the probability that for a population size $N$ there will be no detections is $(1-0.68)^N$. For $N=2$ the likelihood is only $\sim 10\%$; for $N=4$ it has fallen to $\approx 1\%$. Thus, the fact that we did not observe any genuine clusters in the target sample suggests that there is almost certainly not more than $\sim 4$ Local Group IGCs with $M_V < -6$ to be found across the SDSS footprint.  Of course, it is possible to hide any number of objects fainter than this because the \citet{dtzz:15} algorithm cannot find them; if we continue to base our assumptions on the outer halo of M31 then we might expect roughly one cluster fainter than $M_V = -6$ for every two clusters brighter than this level\footnote{Although the PAndAS sample starts to become incomplete below $M_V \approx -6$, with $50\%$ completeness at $M_V \sim -4.1$ \citep{huxor:14}.}.

It is difficult to scale these limits to the whole sky because we do not know how IGCs might be distributed within the Local Group. The simplest assumption of an isotropic distribution on the sky leads to a scale factor $\approx 3$; however the distribution is almost certainly not isotropic due to our vantage point away from the Local Group barycentre, and because the processes that might form IGCs very likely do not distribute them uniformly in any case -- strong galaxy-galaxy interactions tend to create streams and arcs, while small dark matter haloes tend to cluster around larger dark matter haloes. Thus, while it is improbable that a substantial population of relatively luminous IGCs might remain undetected in the SDSS footprint, it would still be very worthwhile to execute similar searches to that conducted by \citet{dtzz:15} on extant and future large-area data sets covering different regions (e.g., those from PS1, DES, SkyMapper, LSST, etc).

\section*{Acknowledgments}
The authors are grateful to the organisers of the Lorentz Center workshop {\it Globular Clusters and Galaxy Halos} in February 2016, where this collaboration was established. ADM acknowledges support from the Australian Research Council through Discovery Projects DP1093431, DP120101237, and DP150103294. MAB was supported by the Programa Nacional de Astronom\'ia y Astrof\'isica from the Spanish Ministry of Economy and Competitiveness, under grants  AYA2013-48226-C3-1-P, AYA2013-48226-C3-2-P, AYA2013-48226-C3-3-P. RL acknowledges funding from the Natural Sciences and Engineering Research Council of Canada PDF award.

This paper is based on observations obtained at the Gemini Observatory, which is operated by the Association of Universities for Research in Astronomy, Inc., under a cooperative agreement with the NSF on behalf of the Gemini partnership: the National Science Foundation (United States), the National Research Council (Canada), CONICYT (Chile), the Australian Research Council (Australia), Minist\'{e}rio da Ci\^{e}ncia, Tecnologia e Inova\c{c}\~{a}o (Brazil) and Ministerio de Ciencia, Tecnolog\'{i}a e Innovaci\'{o}n Productiva (Argentina). These observations were obtained under programmes GN-2008B-Q-22, GN-2014B-Q-26, and GN-2015B-Q-17.





\begin{thebibliography}{99}
\bibitem[\protect\citeauthoryear{Bellazzini et al.}{2003}]{bellazzini:03}
  Bellazzini M., Ibata R., Ferraro F.~R., Testa V., 2003, A\&A, 405, 577
\bibitem[\protect\citeauthoryear{Belokurov et al.}{2006}]{belokurov:06}
  Belokurov V., et al., 2006, ApJ, 642, L137
\bibitem[\protect\citeauthoryear{Belokurov et al.}{2014}]{belokurov:14}
  Belokurov V., Irwin M.~J., Koposov S.~E., Evans N.~W., Gonzalez-Solares E., Metcalfe N., Shanks T., 2014, MNRAS, 441, 2124
\bibitem[\protect\citeauthoryear{Besla et al.}{2016}]{besla:16}
  Besla G., Mart\'{i}nez-Delgado D., van der Marel R.~P., Beletsky Y., Seibert M., Schlafly E.~F., Grebel E.~K., Neyer F., 2016, ApJ, submitted (arXiv:1602.04222)
\bibitem[\protect\citeauthoryear{Conn et al.}{2012}]{conn:12}
  Conn A.~R., et al., 2012, ApJ, 758, 11
\bibitem[\protect\citeauthoryear{Conroy, Loeb \& Spergel}{2011}]{conroy:11}
  Conroy C., Loeb A., Spergel D.~N., 2011, ApJ, 741, 72
\bibitem[\protect\citeauthoryear{Crnojevi\'{c} et al.}{2014}]{crnojevic:14}
  Crnojevi\'{c} D., et al., 2014, MNRAS, 445, 3862
\bibitem[\protect\citeauthoryear{di Tullio Zinn \& Zinn}{2013}]{dtzz:13}
  di Tullio Zinn G., Zinn R., 2013, AJ, 145, 50
\bibitem[\protect\citeauthoryear{di Tullio Zinn \& Zinn}{2014}]{dtzz:14}
  di Tullio Zinn G., Zinn R., 2014, AJ, 147, 90
\bibitem[\protect\citeauthoryear{di Tullio Zinn \& Zinn}{2015}]{dtzz:15}
  di Tullio Zinn G., Zinn R., 2015, AJ, 149, 139
\bibitem[\protect\citeauthoryear{Fardal et al.}{2013}]{fardal:13}
  Fardal M.~A., et al., 2013, MNRAS, 434, 2779
\bibitem[\protect\citeauthoryear{Gregg et al.}{2009}]{gregg:09}
  Gregg M.~D., et al., 2009, AJ, 137, 498
\bibitem[\protect\citeauthoryear{Grillmair \& Carlin}{2016}]{grillmair:16}
  Grillmair C.~J., Carlin J.~L., 2016, in Astrophysics and Space Science Library Vol. 420, {\it Tidal Streams in the Local Group and Beyond}, Springer International Publishing Switzerland, p.87
\bibitem[\protect\citeauthoryear{Huxor et al.}{2008}]{huxor:08} 
  Huxor A.~P., Tanvir N.~R., Ferguson A.~M.~N., Irwin M.~J., Ibata R., Bridges T., Lewis G., 2008, MNRAS, 385, 1989 
\bibitem[\protect\citeauthoryear{Huxor et al.}{2014}]{huxor:14}
  Huxor A.~P., et al., 2014, MNRAS, 442, 2165
\bibitem[\protect\citeauthoryear{Ibata, Gilmore \& Irwin}{1994}]{ibata:94}
  Ibata R.~A., Gilmore G.~F., Irwin M.~J., 1994, Nature, 370, 194
\bibitem[\protect\citeauthoryear{Ibata et al.}{2013}]{ibata:13}
  Ibata R.~A., Nipoti C., Sollima A., Bellazzini M., Chapman S.~C., Dalessandro E., 2013, MNRAS, 428, 3648
\bibitem[\protect\citeauthoryear{Ibata et al.}{2014}]{ibata:14} 
  Ibata R.~A., et al., 2014, ApJ, 780, 128
\bibitem[\protect\citeauthoryear{Laevens et al.}{2014}]{laevens:14}
  Laevens B.~P., et al., 2014, ApJ, 786, L3
\bibitem[\protect\citeauthoryear{Laevens et al.}{2015}]{laevens:15}
  Laevens B.~P., et al., 2015, ApJ, 813, 44
\bibitem[\protect\citeauthoryear{Lane et al.}{2010}]{lane:10}
  Lane R.~R., et al., 2010, MNRAS, 406, 2732
\bibitem[\protect\citeauthoryear{Law \& Majewski}{2010}]{law:10}
  Law D.~R., Majewski S.~R., 2010, ApJ, 718, 1128
\bibitem[\protect\citeauthoryear{Mackey et al.}{2010a}]{mackey:10a} 
  Mackey A.~D., et al., 2010a, ApJ, 717, L11
\bibitem[\protect\citeauthoryear{Mackey et al.}{2010b}]{mackey:10b}
  Mackey A.~D., et al., 2010b, MNRAS, 401, 533
\bibitem[\protect\citeauthoryear{Mackey et al.}{2013}]{mackey:13} 
  Mackey A.~D., et al., 2013, MNRAS, 429, 281
\bibitem[\protect\citeauthoryear{Mackey et al.}{2014}]{mackey:14}
  Mackey A.~D., et al., 2014, MNRAS, 445, L89
\bibitem[\protect\citeauthoryear{Mackey et al.}{2016}]{mackey:16}
  Mackey A.~D., Koposov S.~E., Erkal D., Belokurov V., Da Costa G.~S., G\'{o}mez F.~A., 2016, MNRAS, in press (arXiv:1508.01356)
\bibitem[\protect\citeauthoryear{Majewski et al.}{2003}]{majewski:03}
  Majewski S.~R., Skrutskie M.~F., Weinberg M.~D., Ostheimer J.~C., 2003, ApJ, 599, 1082
\bibitem[\protect\citeauthoryear{Martin et al.}{2014}]{martin:14}
  Martin N.~F., et al., 2014, ApJ, 787, 19
\bibitem[\protect\citeauthoryear{Mashchenko \& Sills}{2005}]{mashchenko:05}
  Mashchenko S., Sills A., 2005, ApJ, 619, 258
\bibitem[\protect\citeauthoryear{McConnachie et al.}{2009}]{mcconnachie:09} 
  McConnachie A.~W., et al., 2009, Nature, 461, 66 
\bibitem[\protect\citeauthoryear{McConnachie et al.}{2010}]{mcconnachie:10} 
  McConnachie A.~W., Ferguson A.~M.~N., Irwin M.~J., Dubinski J., Widrow L.~M., Dotter A., Ibata R., Lewis G.~F., 2010, ApJ, 723, 1038
\bibitem[\protect\citeauthoryear{McConnachie}{2012}]{mcconnachie:12} 
  McConnachie A.~W., 2012, AJ, 144, 4
\bibitem[\protect\citeauthoryear{Peebles}{1984}]{peebles:84}
  Peebles P.~J.~E., 1984, ApJ, 277, 470
\bibitem[\protect\citeauthoryear{Peng et al.}{2011}]{peng:11}
  Peng E.~W., et al., 2011, ApJ, 730, 23
\bibitem[\protect\citeauthoryear{Samsing}{2015}]{samsing:15}
  Samsing J., 2015, ApJ, 799, 145
\bibitem[\protect\citeauthoryear{Veljanoski et al.}{2013}]{veljanoski:13} 
  Veljanoski J., et al., 2013, ApJ, 768, L33
\bibitem[\protect\citeauthoryear{Veljanoski et al.}{2014}]{veljanoski:14}
  Veljanoski J., et al., 2014, MNRAS, 442, 2929
\bibitem[\protect\citeauthoryear{West et al.}{2011}]{west:11}
  West M.~J., Jord\'{a}n A., Blakeslee J.~P., C\^{o}t\'{e} P., Gregg M.~D., Takamiya M., Marzke R.~O., 2011, A\&A, 528, 115
\end{thebibliography}







\bsp	
\label{lastpage}
\end{document}